\documentclass[aps,prd,onecolumn,groupedaddress,showpacs,nofootinbib,amssymb]{revtex4}
\usepackage[dvips]{graphicx}
\usepackage{amssymb}
\usepackage{amsmath}
\usepackage{graphicx}
\usepackage{amsfonts}
\usepackage{bm}
\usepackage{amsmath,amssymb,theorem}

\begin{document}

\title{Autonomous Dynamical System Description of de Sitter Evolution in Scalar Assisted $f(R)-\phi$ Gravity}
\author{
K. Kleidis,$^{1}$\,\thanks{kleidis@teiser.gr}
V.K. Oikonomou$^{2,1}$,\,\thanks{v.k.oikonomou1979@gmail.com}}
\affiliation{
$^{1)}$ Department of Mechanical Engineering\\ Technological
Education Institute of Central Macedonia \\
62124 Serres, Greece \\
$^{2)}$ Department of Physics, Aristotle University of Thessaloniki, Thessaloniki 54124, Greece
}

\tolerance=5000

\begin{abstract}
In this letter we will study the cosmological dynamical system of an
$f(R)$ gravity in the presence of a canonical scalar field $\phi$
with an exponential potential, by constructing the dynamical system
in a way that it is render autonomous. This feature is controlled by
a single variable $m$, which when it is constant, the dynamical
system is autonomous. We focus on the $m=0$ case which, as we
demonstrate by using a numerical analysis approach, leads to an
unstable de Sitter attractor, which occurs after $N\sim 60$
$e$-foldings. This instability  can be viewed as a graceful exit
from inflation, which is inherent to the dynamics of de Sitter
attractors.
\end{abstract}


\maketitle

\section{Introduction}

The phase space structure of a cosmological dynamical system may
reveal important information regarding the existence and stability
of the fixed points. Also the existence of a fixed point with
physical significance may can also be revealed by the study of the
cosmological dynamical system. In the literature, the dynamical
systems approach has been adopted in various theoretical contexts
\cite{Boehmer:2014vea,Bohmer:2010re,Goheer:2007wu,Leon:2014yua,Guo:2013swa,Leon:2010pu,deSouza:2007zpn,Giacomini:2017yuk,Kofinas:2014aka,Leon:2012mt,Gonzalez:2006cj,Alho:2016gzi,Biswas:2015cva,Muller:2014qja,Mirza:2014nfa,Rippl:1995bg,Ivanov:2011vy,Khurshudyan:2016qox,Boko:2016mwr,Odintsov:2017icc,Granda:2017dlx,Landim:2016gpz,Landim:2015uda,Landim:2016dxh,Bari:2018edl,Chakraborty:2018bxh,Ganiou:2018dta,Shah:2018qkh,Odintsov:2018uaw,Odintsov:2018awm,Oikonomou:2017ppp,Odintsov:2017tbc},
see also \cite{Odintsov:2015wwp}, and all the studies aimed to
reveal the stability structure and the interconnection of fixed
points by using various phase space trajectories. In this letter
we aim to study the behavior of the de Sitter attractors of an
$f(R)$ gravity \cite{reviews1,reviews2,reviews3,reviews4}, in the
presence of a canonical scalar field with potential $V(\phi)$,
which we denote for simplicity as $f(R)-\phi$ gravity. For a
recent inflationary realization in the context of $f(R)-\phi$
gravity, see Ref. \cite{Kleidis:2018fdu}. We will construct an
autonomous dynamical system by appropriately manipulating the
cosmological equations, and we will study the specific behavior of
de Sitter attractors. This work is in the spirit of Refs.
\cite{Oikonomou:2017ppp,Odintsov:2017tbc}, where an autonomous
dynamical system approach was also considered. As we will show,
the dynamical system is rendered autonomous only when the scalar
potential is an exponential \cite{new}, so by making this
assumption, we will investigate the behavior of the de Sitter
attractors in this particular case. The importance of having an
autonomous dynamical system can be revealed by the following
example, which can be found in Ref. \cite{dynsystemsbook},
$\dot{x}=-x+t$, which can be solved and the solution is
$x(t)=t-1+e^{-t}(x_0+1)$. As it is obvious, the solution
approaches $t-1$ asymptotically, and also the fixed point is
$x=t$. However the fixed point $x=t$ is not a solution of the
dynamical system, and if the standard theorems for fixed point are
applied, the wrong conclusion of having the solutions approaching
$x=t$ is obtained. In view of the above example, it is conceivable
that an autonomous cosmological dynamical system may reveal
important information with regard the existence and stability of
fixed points. As we will demonstrate in this letter, the only
parameter that contains a time-dependence in the dynamical system
is $m=-\frac{\ddot{H}}{H^3}$, with $H$ being the Hubble rate of
the Universe. Hence we will assume that this parameter is equal to
zero, without specifying the cosmological evolution for which this
is possible. As it turns out, there exists a fixed point which is
actually a de Sitter fixed point. By carefully examining the phase
structure, we will demonstrate by using a numerical analysis, that
this fixed point is reached for $N\sim 50-60$ $e$-foldings,
however at $N\sim 70$ it is rendered unstable due to an existing
instability in some variables. Therefore, this actually describes
a de Sitter attractor which becomes internally unstable, a feature
that can be viewed as an exit from the inflationary attractor
solution.

This letter is organized as follows: In section II, we present the
general structure of the autonomous dynamical system, and we
demonstrate why the exponential potential is needed in order to
render the dynamical system autonomous. In section III we perform an
in depth numerical analysis, in which we investigate the existence
and the stability of the de Sitter attractor. Also we examine the
qualitative features of the phase space, which reveals whether the
de Sitter fixed point is stable or not. Finally, the conclusions
follow in the end of the letter.

Before we proceed, let us fix the geometric background, which is a flat
Friedmann-Robertson-Walker (FRW) metric, with line element,
\begin{equation}\label{frw}
ds^2 = - dt^2 + a(t)^2 \sum_{i=1,2,3} \left(dx^i\right)^2\, ,
\end{equation}
with $a(t)$  being the scale factor. Also for the FRW metric, the Ricci
scalar is equal to,
\begin{equation}\label{ricciscalaranalytic}
R=6\left (\dot{H}+2H^2 \right )\, ,
\end{equation}
with $H=\frac{\dot{a}}{a}$, being as usual the Hubble rate.


\section{The Autonomous Dynamical System of $f(R,\phi)$ Gravity with Exponential Potential}

As we briefly mentioned in the introduction, having the cosmological
evolution expressed in terms of an autonomous dynamical system, may
be vital for the correct description of the dynamical evolution. So
in this section, we shall present the general structure of an
$f(R)$-$\phi$ scalar-tensor gravity and we shall use the equations
of motion in order to obtain an autonomous dynamical system. A
general $f(R)$-$\phi$ gravity action, with scalar potential
$V(\phi)$ has the following form,
\begin{equation}\label{action}
\mathcal{S}=\int \mathrm{d}^4x\sqrt{-g}\left(
\frac{f(R)}{2\kappa^2}-\frac{1}{2}\partial_{\mu}\phi\partial^{\mu}\phi-V(\phi)
\right)\, ,
\end{equation}
where $\kappa^2=8\pi G=\frac{1}{M_p^2}$ and $M_p$ stands for the Planck
mass scale. By varying the action with respect to the metric, we obtain,
\begin{equation}\label{eqnmotion}
F(R)R_{\mu \nu}(g)-\frac{1}{2}f(R)g_{\mu
\nu}-\nabla_{\mu}\nabla_{\nu}f(R)+g_{\mu \nu}\square
F(R)=\kappa^2\left (\nabla_{\mu}\phi \nabla_{\nu}\phi
\right)-\frac{1}{2}g_{\mu \nu}\nabla^{\rho}\nabla_{\rho}-Vg_{\mu
\nu}\, ,
\end{equation}
and also the scalar field satisfies,
\begin{equation}\label{newentry}
\square\phi=V'(\phi)\, ,
\end{equation}
where the prime indicates differentiation with respect to the scalar
field $\phi$. For the metric (\ref{frw}), and by assuming that the
scalar field depends solely on the cosmic time, the cosmological
equations can be written in the following form,
\begin{align}
\label{JGRG15} & 3FH^2+\frac{1}{2}\left(
f-RF\right)+3H\dot{F}=\kappa^2\left(
\frac{1}{2}\dot{\phi}^2+V(\phi)\right),\\ \notag &
-2F\dot{H}-\ddot{F}+H\dot{F}=\kappa^2\dot{\phi}^2, \\ \notag &
\ddot{\phi}+3H\dot{\phi}+V'=0 \, ,
\end{align}
where the ``dot'' indicates differentiation with respect to the
cosmic time. The autonomous dynamical system of $f(R)-\phi$ gravity
can be obtained if we use the following variables,
\begin{equation}\label{variablesslowdown}
x_1=-\frac{\dot{F}(R)}{F(R)H},\,\,\,x_2=-\frac{f(R)}{6F(R)H^2},\,\,\,x_3=
\frac{R}{6H^2},\,\,\,x_4=\frac{V(\kappa^2\phi)}{6FH^2},\,\,\,x_5=\frac{\kappa^2\dot{\phi}^2}{6FH^2},\,\,\, x_6=\frac{\kappa^2}{6F}\, .
\end{equation}
The case with a general potential $V(\phi)$ is difficult to tackle,
since in all cases an non-autonomous system is constructed, so we
focus on the case that the potential is equal to
$V(\phi)=e^{-\lambda \phi}$.

A convenient variable that may quantify in an optimal way the duration of de Sitter phase is the $e$-foldings number $N$, so by using the variables
(\ref{variablesslowdown}) and also the equations (\ref{JGRG15}), we get the following dynamical
system,
\begin{align}\label{dynamicalsystemmain}
& \frac{\mathrm{d}x_1}{\mathrm{d}N}=-x_1^2-x_1 x_3-3 x_1+2 x_3+6
x_5-4\, ,
\\ \notag &
\frac{\mathrm{d}x_2}{\mathrm{d}N}=m+x_1 x_2-2 x_2 x_3+4 x_2-4 x_3+8\, ,\\
\notag & \frac{\mathrm{d}x_3}{\mathrm{d}N}=-m-2 x_3 x_3+8 x_3-8 \, ,
\\ \notag & \frac{\mathrm{d}x_4}{\mathrm{d}N}=x_1
x_4-\frac{x_3 x_4}{3}+\frac{\lambda  x_4
\sqrt{x_5}}{\sqrt{x_6}}+\frac{2 x_4}{3} \, , \\
\notag & \frac{\mathrm{d}x_5}{\mathrm{d}N}=x_1 x_4-2 x_3 x_5+\frac{2
\lambda  x_4 \sqrt{x_5}}{\sqrt{x_6}}+4 x_5 \, , \\
\notag & \frac{\mathrm{d}x_6}{\mathrm{d}N}=x_1 x_6 \, .
\end{align}
where the parameter $m$ stands for,
\begin{equation}\label{parameterm}
m=-\frac{\ddot{H}}{H^3}\, .
\end{equation}
Since we are interested in de Sitter or quasi-de Sitter attractors,
the variable $m$ for a quasi-de Sitter attractor is zero. So in
order to reveal the structure of the quasi de Sitter attractor, we
shall focus on the case of $m=0$ and we shall perform an analysis of
the behavior of the dynamical system (\ref{dynamicalsystemmain}).
The effective equation of state (EoS) for a general $f(R,\phi)$
theory is equal to,
\begin{equation}\label{weffoneeqn}
w_{eff}=-1-\frac{2\dot{H}}{3H^2}\, ,
\end{equation}
and by using the variable $x_3$, it can be expressed in terms of $x_3$ as follows,
\begin{equation}\label{eos1}
w_{eff}=-\frac{1}{3} (2 x_3-1)\, .
\end{equation}
In the following we shall focus on the behavior of the de Sitter and
of the quasi-de Sitter attractors.
\begin{figure}[h]
\centering
\includegraphics[width=20pc]{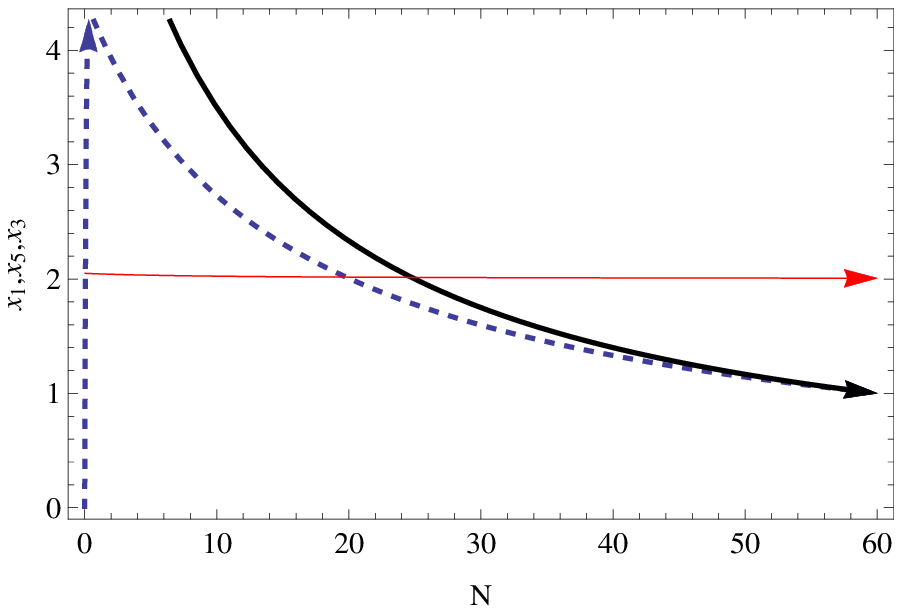}
\includegraphics[width=20pc]{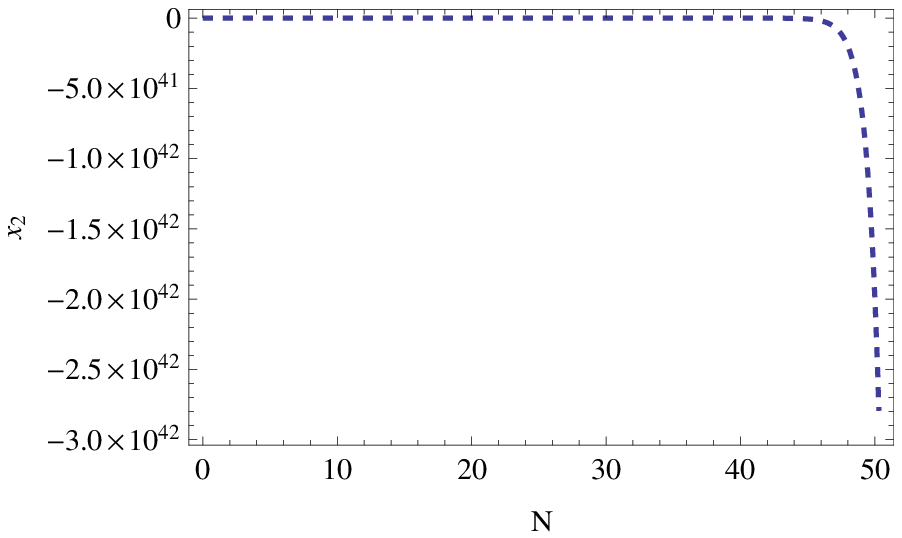}
\includegraphics[width=20pc]{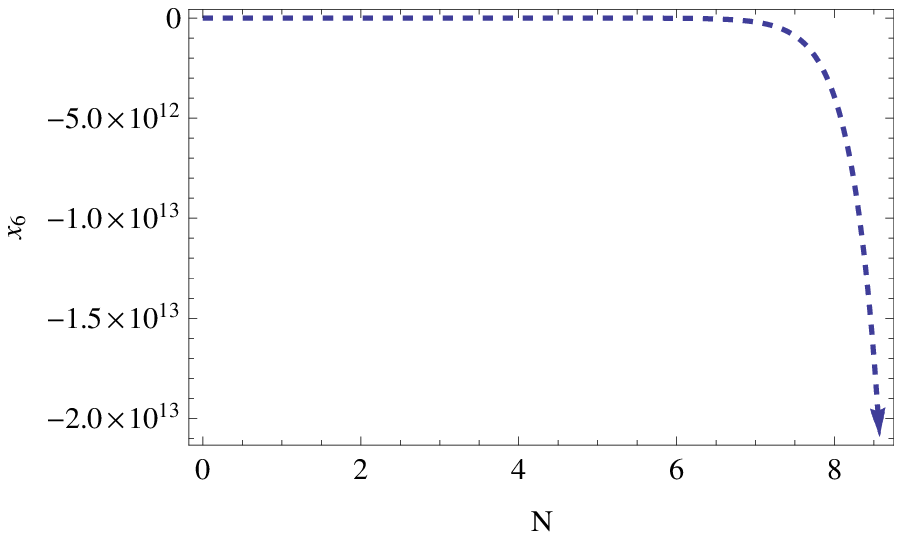}
\caption{{\it{Numerical solutions $x_1(N)$, $x_2(N)$, $x_3(N)$,
$x_4(N)$, $x_5(N)$ and $x_6(N)$ for the dynamical system
(\ref{dynamicalsystemmain}), for the initial conditions
$x_1(0)=-0.01$, $x_2(0)=0$, $x_3(0)=2.05$, $x_4(0)=0$, $x_5(0)=7$,
$x_6(0)=-2$,  and for $m=0$, $\lambda=100$. In the upper left, the
black curve corresponds to the variable $x_5$, the red curve to the
parameter $x_3$ and the blue curve to the parameter $x_1$.}}}
\label{plot1}
\end{figure}

\subsection{Study of the Phase Space and the de Sitter Attractor}

As we already stressed earlier, the only source of non-autonomous
structure in the dynamical system is contained in the parameter $m$,
and for certain cosmological evolutions this parameter is constant.
For example, if we choose $m=0$, this can occur for a quasi-de
Sitter evolution with scale factor $a(t)=e^{H_0 t-H_i t^2}$, but
also the symmetric bounce yields $m=0$, in which case the scale
factor is $a(t)=e^{A\, t^2}$.
\begin{figure}[h]
\centering
\includegraphics[width=20pc]{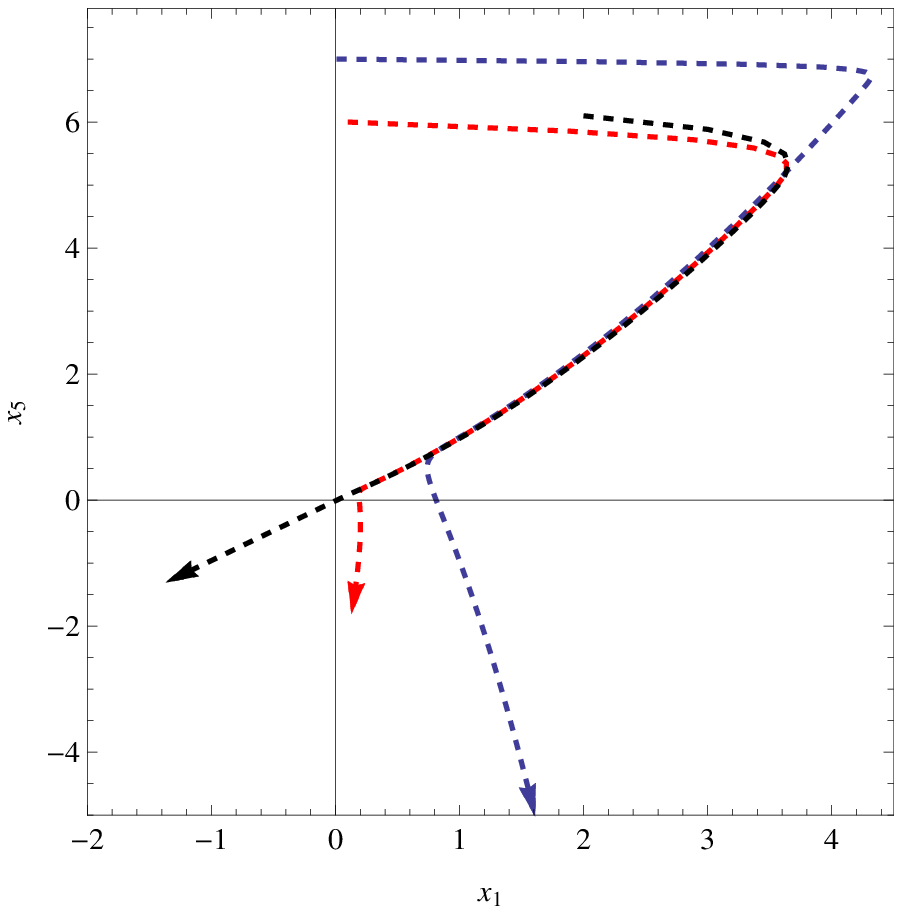}
\includegraphics[width=20pc]{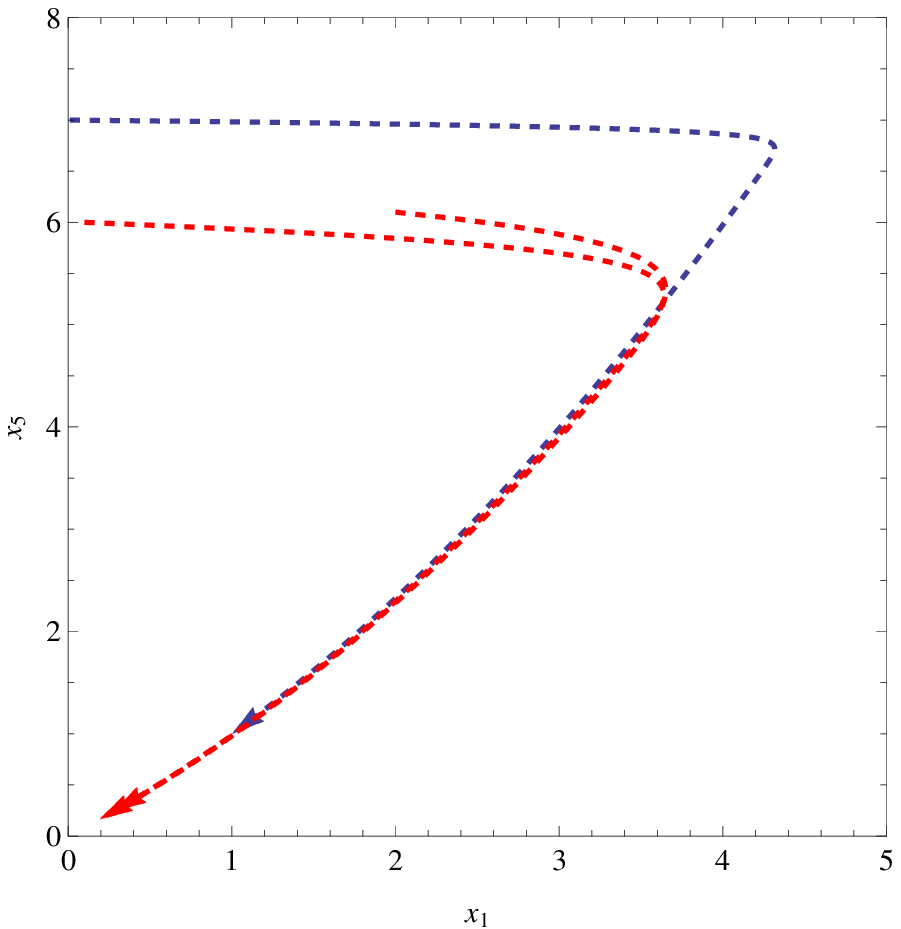}
\caption{{\it{Parametric plot in the plane $x_1-x_5$ for $N=(0,60)$
(right plot), and $N=(0,80)$ (left plot) for the dynamical system
(\ref{dynamicalsystemmain}),  with $m=0$, $\lambda=100$, and for
various initial conditions.}}} \label{plot2}
\end{figure}
Also the case $m=-9/2$ can occur for a matter dominated cosmological
evolution with $a(t)\sim t^{2/3}$. In this work we shall focus on
the case $m=0$, however without specifying the specific choice of
the Hubble rate, and also note that the general form of the $f(R)$
gravity is not specified. The fixed points of the dynamical system
(\ref{dynamicalsystemmain}) can be found by solving the system of
equations $f_i=0$, $i=1,..,6$, and the only solution is
$\phi_1^*=(x_1,x_3,x_5)=(0,2,0)$. The rest of the variables are not
fixed, and in the rest of this section we shall investigate the
behavior all the variables. Firstly, let us note that the case
$x_3=2$ corresponds to $w_{eff}=-1$, as can be verified by Eq.
(\ref{eos1}), so the case $m=0$ yields a fixed point with
$w_{eff}=-1$, and therefore this is a de Sitter fixed point. The
stability of this point cannot be addressed in a conventional way,
as it can be shown, since the linearization matrix contains entries
which are infinite. So we shall numerically solve the system of
differential equations (\ref{dynamicalsystemmain}) by using various
initial conditions, and for various values of the $e$-foldings
number $N$. As we now demonstrate, the results are particularly
interesting, since once the de Sitter attractor is reached around
$N\sim 60$ $e$-foldings, the de Sitter attractor becomes unstable.
We have solved numerically the system of differential equations, for
the initial conditions $x_1(0)=-0.01$, $x_2(0)=0$, $x_3(0)=2.05$,
$x_4(0)=0$, $x_5(0)=7$, $x_6(0)=-2$, and in Fig. \ref{plot1} we plot
the behavior of the parameters $x_i$ as functions of the
$e$-foldings number $N$. In the upper left, the black curve
corresponds to the variable $x_5$, the red curve to the parameter
$x_3$ and the blue curve to the parameter $x_1$. As it can be seen,
the variable $x_3$ tends to $x_3=2$ quite fast, and for $N<60$,
however the fixed point values for $x_1$ and $x_5$ are not reached
at all. Therefore, this shows that although the de Sitter attractor
is reached, it is unstable due to the behavior of the variables
$x_1$ and $x_5$.
\begin{figure}[h]
\centering
\includegraphics[width=20pc]{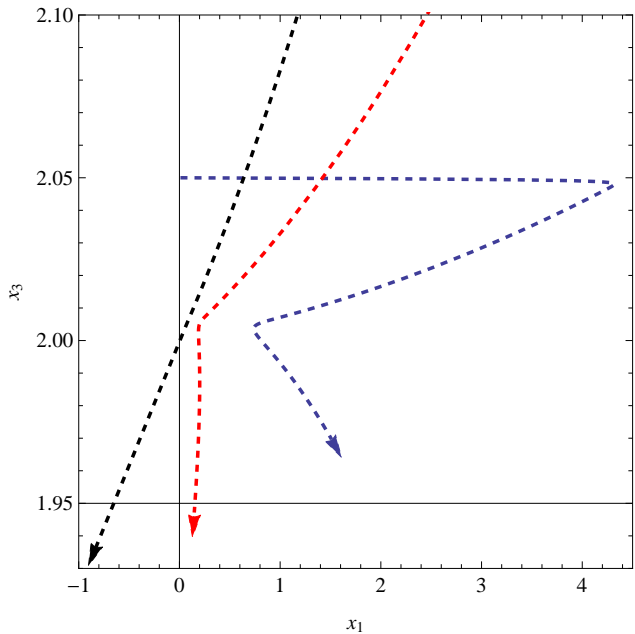}
\caption{{\it{Parametric plot in the plane $x_1-x_3$ for $N=(0,80)$
for the dynamical system (\ref{dynamicalsystemmain}),  with $m=0$,
$\lambda=100$, and for various initial conditions.}}} \label{plot3}
\end{figure}
The instability in the variables $x_1$ and $x_5$ of the fixed point
values $(x_1,x_5)=(0,0)$ can also be seen in the $x_1-x_5$ plane,
and we present the plot in Fig. \ref{plot2}, for various initial
conditions. As it can be seen, the fixed point values are approached
around $N\sim 60$, and the three curves meet each other, however,
around $N\sim 80-90$, the three curves split. This clearly indicates
that the de Sitter attractor $x_3=2$, becomes eventually unstable
after $N\sim 70$ $e$-foldings. The same can also be seen in the
$x_1-x_3$ plane, which we plot in Fig. \ref{plot3}, where although
in all the curves, the value $x_3$ is reached, the curves split and
the attractor $x_3=2$ becomes eventually unstable.

Let us briefly discuss the physical interpretation of the present
numerical analysis. As it seems from Figs. \ref{plot1}, \ref{plot2}
and \ref{plot3}, the de Sitter attractor $x_3=2$ is reached for
$N\sim 60$, however, due to the fact that the variables $x_1$ and
$x_5$ become unstable, this de Sitter attractor is rendered
unstable. Hence, if the de Sitter attractor is an inflationary
attractor, then although the attractor $x_3=2$ is reached for $N\sim
60$ and even earlier as it can be seen in the upper left plot of
Fig. \ref{plot1}, the de Sitter attractor becomes unstable, and this
can be seen as a graceful exit from the inflationary era.

Hence the $f(R)-\phi$ theory with exponential scalar potential, can
describe accurately an inflationary era which comes to an end
eventually, due to inherent instabilities. We have also checked the
case $m=-9/2$ and we found that the matter dominated era cannot be
consistently described by a $f(R)-\phi$, so perfect fluids are
needed to describe this era, however we defer from going into
details for brevity.

\section{Conclusions}

In this letter we investigated the dynamical evolution of the
autonomous system which is constructed from an exponential
$f(R)-\phi$ gravity. By appropriately choosing the variables related
to the physical quantities entering the equations of motion, we were
able to construct an dynamical system, in which the only time
dependence is solely contained in the variable
$m=-\frac{\ddot{H}}{H^3}$. We focused on the specific case $m=0$,
without specifying the Hubble rate and we investigated the behavior
of the corresponding autonomous dynamical system. As we
demonstrated, the case $m=0$ leads to a de Sitter attractor
solution, and by employing a numerical analysis investigation, we
demonstrated that the de Sitter fixed point is reached at $N\sim
50-60$ $e$-foldings. After $N\sim 60$ some variables of the
dynamical system become unstable, and as a result, the dynamical
system is rendered unstable and the de Sitter attractor ceases to be
the final attractor of the theory. As we argued, this feature
resembles the graceful exit from the inflationary de Sitter
attractor, so this seems to be the case. In effect, the exponential
$f(R)-\phi$ gravity has a de Sitter attractor which is reached
around $N\sim 50-60$ and after that it becomes unstable. This
instability could be viewed as a graceful exit from the inflationary
de Sitter attractor, and this graceful exit is inherent to the
dynamics of de Sitter attractors.

We need to note that the instability of the de Sitter attractor is
not guaranteed for all theories of modified gravity which are
studied in the way we just presented. Actually, a pure $f(R)$
gravity theory, even in the presence of matter and radiation perfect
fluids, has a stable de Sitter attractor corresponding to the case
$m=0$. This study will be reported elsewhere.

\section*{Acknowledgments}

Financial support by the Research
Committee of the Technological Education Institute of Central Macedonia, Serres,
under grant SAT/ME/230518-126/15, is gratefully acknowledged.

\end{document}